%% file: Motion_adapted_3D_FSE_ICASSP_arxiv.tex
\documentclass{article}

\usepackage[preprint]{spconf}
\usepackage{amsmath,graphicx,color,hhline}
\usepackage[hidelinks]{hyperref}

\copyrightnotice{\fbox{\parbox{\dimexpr\textwidth-\fboxsep-\fboxrule\relax}{\footnotesize \textcopyright 2019 IEEE. Personal use of this material is permitted. Permission from IEEE must be obtained for all other uses, in any current or future media, including reprinting/republishing this material for advertising or promotional purposes, creating new collective works, for resale or redistribution to servers or lists, or reuse of any copyrighted component of this work in other works. DOI: \href{https://doi.org/10.1109/ICASSP.2019.8683218}{10.1109/ICASSP.2019.868321}
}}}

\title{Motion-Adapted Three-Dimensional Frequency Selective Extrapolation}
\name{Andreas Spruck, Markus Jonscher, J\"urgen Seiler, and Andr\'e Kaup}
\address{Friedrich-Alexander Universit\"at Erlangen-N\"urnberg (FAU)\\
	Multimedia Communications and Signal Processing\\
	Cauerstr. 7, 91058 Erlangen, Germany}

\begin{document}
%\ninept
%
\maketitle
\begin{abstract}
It has been shown, that high resolution images can be acquired using a low resolution sensor with non-regular sampling. Therefore, post-processing is necessary. In terms of video data, not only the spatial neighborhood can be used to assist the reconstruction, but also the temporal neighborhood. A popular and well performing algorithm for this kind of problem is the three-dimensional frequency selective extrapolation (3D-FSE) for which a motion adapted version is introduced in this paper. This proposed extension solves the problem of changing content within the area considered by the 3D-FSE, which is caused by motion within the sequence. Because of this motion, it may happen that regions are emphasized during the reconstruction that are not present in the original signal within the considered area. By that, false content is introduced into the extrapolated sequence, which affects the resulting image quality negatively. The novel extension, presented in the following, incorporates motion data of the sequence in order to adapt the algorithm accordingly, and compensates changing content, resulting in gains of up to 1.75 dB compared to the existing 3D-FSE.
\end{abstract}
\section{Introduction}
\label{sec:intro}
The demand for algorithms that are capable of generating high resolution video data from low resolution data grows steadily. The fields of application for such algorithms are manifold and reach from medical imaging over security surveillance to consumer applications like digital remastering of video data. Due to this, there are many different approaches towards this problem. A short overview of the most common methods can be seen in \cite{Park2003}.
Another possible method to obtain high resolution image or video data, on which we focus here, is to use a low resolution non-regular sampling sensor for acquisition and applying a reconstruction algorithm on the recorded data afterwards, to recover the image areas where no samples were captured. This method was first introduced in \cite{Schoeberl2011} where the two-dimensional frequency selective extrapolation (2D-FSE) \cite{seiler2010complex} was used to extrapolate the non-available areas. As we consider video sequences here, the three-dimensional frequency selective extrapolation (3D-FSE) \cite{Meisinger2007a} is used to reconstruct the not directly acquired areas, as it also incorporates the temporal component of the sequence. Moreover, this kind of sampling bears the benefit that aliasing artifacts can be reduced in a visible noticeable amount as shown in \cite{Hennenfent2007,Maeda2009}. 

The paper is structured as follows, in Section~\ref{sec:nonregrec} the basic principle of non-regular sampling will be shown. Section~\ref{sec:3D-FSR} introduces the three-dimensional frequency selective extrapolation. The novel motion adaptive extension will be explained in further detail in Section~\ref{sec:mc_weighting}. The simulation results will be presented and discussed in Section~\ref{sec:simulations} before we summarize and conclude our paper in Section~\ref{sec:conclusion}.
\vspace{-0.3cm}
\section{Reconstruction of non-regular sampled data}
\label{sec:nonregrec}
\begin{figure}[t]
	\centering
	\def\svgwidth{0.45\textwidth}
	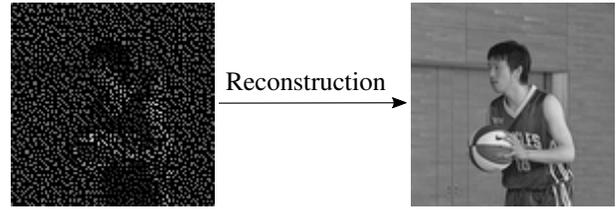
	\caption{Schematic representation of the image acquisition using a non-regular sampling sensor.}
	\label{fig:mask}
\end{figure}
The recording scenario that is applied during this work was originally presented in \cite{Schoeberl2011}. The sensor introduced there is a low resolution image sensor, where only one fourth of the area of every pixel is sensitive to light. The light sensitive areas are non-regularly distributed among the four quadrants of each pixel. As only a small area of the large low resolution pixel is sensitive to light, this active area can also be regarded as a single pixel of a sensor with a four times higher resolution whose pixels are fully light sensitive. 
By describing the sensor in that way, it can be regarded as a high resolution sensor covered with a non-regular mask. % as shown in Figure~\ref{fig:mask}.
This mask follows the aforementioned description and leaves only one pixel out of a $2\times2$ block unmasked. By this a high resolution image can be acquired while only reading out and storing one fourth of its pixels.
As we are processing video data and use the same sensor throughout this work, every frame of the sequence is masked with the same mask in order to obtain a non-regular sub-sampled sequence as presented in \cite{Schoeberl2011}. The non-regular sub-sampled sequence can be written as 
\begin{equation}
s_\mathrm{nr}[x,y,t] = s[x,y,t] \cdot b[x,y,t].
\end{equation}
Thereby, $s[x,y,t]$ denotes the full high resolution video data and $b[x,y,t]$ denotes the sub-sampling mask as described before. 
After masking the frames, it becomes obvious that the masked areas need to be reconstructed in order to obtain a satisfying result, as schematically shown in Figure~\ref{fig:mask}. During this work, 3D-FSE is used to reconstruct the signal $s[x,y,t]$ from $s_\mathrm{nr}[x,y,t]$ following the idea of \cite{Schoeberl2011}. This algorithm will be presented in more detail in the next section.
\vspace{-0.3cm}
\section{Three-dimensional frequency selective extrapolation (3D-FSE)}
\label{sec:3D-FSR}
\begin{figure}[t]
	\centering
	\def\svgwidth{0.45\textwidth}
	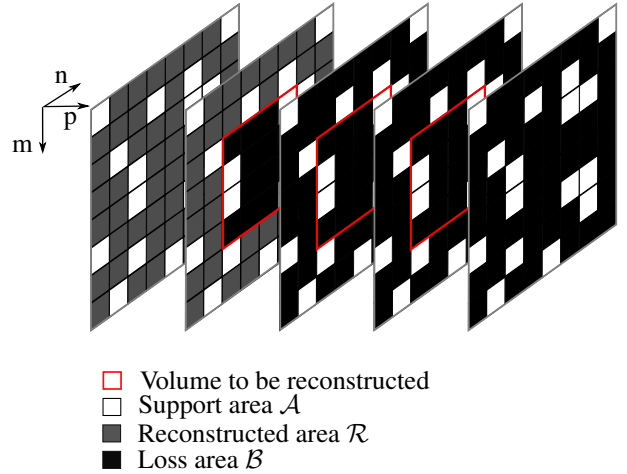
	\caption{Schematic representation of the reconstruction area of the 3D-FSE.}
	\label{fig:block_folge}
\end{figure}

The 3D-FSE used here, is a method to reconstruct lost or defective areas of a video based on the preserved data. 
It is based on 2D-FSE, as described in \cite{seiler2010complex}, and was extended to three dimensions in \cite{Meisinger2007a}. The 3D-FSE reconstructs the sequence in a block-wise manner. Meaning, that every block $f[m,n,p]$ within the sequence is reconstructed separately. Every block $f[m,n,p]$ has thereby a dimension of $M \times N$ pixels in the spatial direction, and a length of $P$ in temporal direction. The basic principle of the algorithm is to iteratively generate a model of superimposed weighted Fourier basis functions, that is used to fill in the missing part of the signal, which is called loss area~$\mathcal{B}$, and is depicted in Figure~\ref{fig:block_folge}. The picture content that is held in the available pixels from the block $f[m,n,p]$ within the non-regular sampled block $f_\mathrm{nr}[m,n,p]$, is called support area~$\mathcal{A}$ and serves as a basis for the model generation. 
In contrast to the 2D-FSE the area that is to be reconstructed extends over several frames, as denoted by the red boxes in {Figure~\ref{fig:block_folge}}. The extrapolation area $\mathcal{L}$ is defined by the variables $m$ and $n$ along the spatial axes and $p$ along the time axis. The extrapolation area $\mathcal{L}$ is composed of the support area $\mathcal{A}$, the loss area $\mathcal{B}$ and the area $\mathcal{R}$, which holds the previously reconstructed values, as $\mathcal{L} = \mathcal{A} \cup \mathcal{B} \cup \mathcal{R}$ as stated in \cite{Meisinger2007a}.

The aim of 3D-FSE is to generate a model which replicates the original signal $f[m,n,p]$ in the volume $\mathcal{L}$. The non-sampled areas of the signal $f_\mathrm{nr}[m,n,p]$ are then replaced by this model. 

The exact description of the algorithm can be found in \cite{Meisinger2007a}. During the scope of this paper, the algorithm from \cite{Meisinger2007a} was extended by using the optimized processing order from \cite{seiler2016optimized}.
By incorporating the weighting function $w[m,n,p]$, it is possible to assign different weights to the different areas of the sequence. By this, it is ensured that missing regions are not taken into account for model generation and pixels far away from the currently considered area are weighted with a smaller weight than areas nearby. The spatial weighting function
\begin{equation}
\begin{aligned}[left]
w[m,n,p]= \begin{cases}
\rho[m,n,p] &, (m,n,p) \in \cal{A}\\
\delta \rho[m,n,p] &, (m,n,p) \in \cal{R}\\
0 &, (m,n,p) \in \cal{B}
\end{cases}
\end{aligned}
\label{equ:weighting}
\end{equation}
used during the selection of the basis function is defined as in \cite{Meisinger2007a}. By varying $\hat{\rho}$ in 
\begin{equation}
\rho[m,n,p] = \hat{\rho}^{\sqrt{\left(m-\frac{M-1}{2}\right)^2+\left(n-\frac{N-1}{2}\right)^2+\left(p-\frac{P-1}{2}\right)^2}} 
\end{equation}
the decay of the weighting function can be adjusted. Parameter $\delta$ controls how much influence previously reconstructed values have on the model generation.

The advantage of the FSE is that the computationally expensive calculations can be performed in the frequency domain as it is shown in \cite{seiler2010complex} and \cite{Meisinger2007a}. By doing this, the generation of the model can be sped up drastically. 
As all operations can be executed in the frequency domain it is sufficient to perform only two transformations, one in the beginning and one in the end \cite{Meisinger2007a}.
\section{Motion compensated weighting}
\label{sec:mc_weighting}
\begin{figure}[t]
	\centering
	\def\svgwidth{0.46\textwidth}
	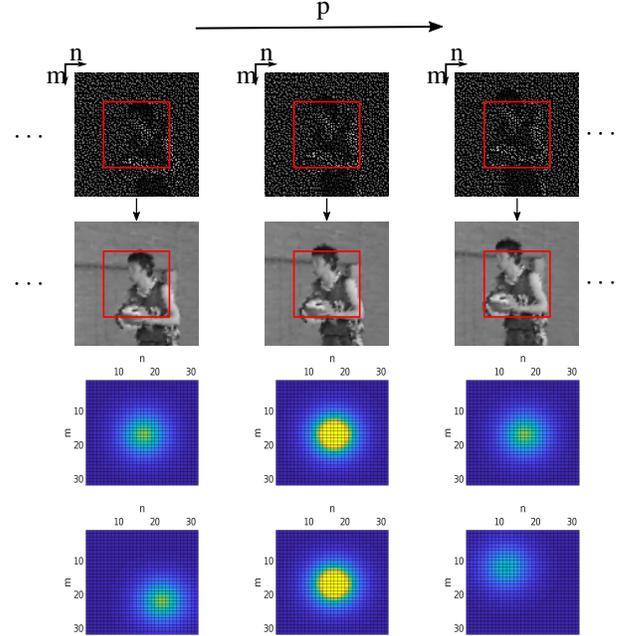
	\caption{First row: Non-regular sampled sequence, second row: linear interpolated sequence, third row: static spatial weighting function for corresponding frame, fourth row: motion compensated spatial weighting function for corresponding frame.}
	\label{fig:overview_FSE}
\end{figure}

The problem occurring with the static weighting function as presented before and used in \cite{Meisinger2007a} and \cite{seiler2016optimized}, is, that due to motion within the sequence, the region of interest at which the maximum of the weighting function is placed, moves out of the maximum of the decaying weighting function over time, as can be seen in {Figure~\ref{fig:overview_FSE}}.
By the fact, that the content of the maximally weighted area varies over time, content that is actually not present in the part of the sequence, that is to be reconstructed, might be weighted with a high weight, resulting in admission of basis functions to the model that are not part of the original signal. This leads to unsharp edges and even ghosting artifacts, resulting in a strong degradation of the overall reconstruction quality. Due to this, the fixed spatial weighting function as presented in (\ref{equ:weighting}) is replaced by a more sophisticated one in this paper.

Different approaches to solving this problem have been made in the past. In \cite{seiler2011}, the extrapolation volume $\mathcal{L}$ was selected, by incorporating motion data, such that the selected area of every frame shows the same image region over time. Thereby it is guaranteed that the weighting function assesses the same regions of the image the maximal weight, as the selection of the support area of the 3D-FSE is performed in a motion compensated manner.
In \cite{jonscher16recursive} another approach is presented. There, a first reconstruction using the two-dimensional frequency selective reconstruction (2D-FSR) from \cite{seiler2015resampling} is performed. After this reconstruction, a motion estimation is performed to search for pixels in the loss area $\mathcal{B}$ of the currently considered frame that are contained in the support area $\mathcal{A}$ in previous or succeeding frames. If such pixels are found, they are copied into the loss area of the current cube. By this, the size of the loss area can be reduced, which results in a far better reconstruction quality of the following 2D-FSR \cite{jonscher16recursive}.

In this paper we pursue a different approach. We shift the weighting function according to the motion in the sequence such that in every frame the maximum of the weighting function is placed over the same content, as it can be seen in the bottom line of {Figure~\ref{fig:overview_FSE}}. 
For the motion estimation, the sub-sampled sequence is reconstructed with a bilinear interpolation in a first step in order to enable a motion estimation. Afterwards a optical flow implementation following \cite{Farnebaeck2003} is used to estimate the motion on the interpolated data. For the estimation of the motion, the whole interpolated frame is regarded. This results in a motion vector field holding a motion vector for every pixel in the frame relative to the currently considered frame. Out of this vector field the cube currently considered by the 3D-FSE is extracted. These vectors are averaged to one overall motion vector for each slice of the cube
\begin{equation}
\begin{pmatrix}
\bar{v}_\mathrm{x}[p]\\ \bar{v}_\mathrm{y}[p]
\end{pmatrix} = \frac{1}{MN} \cdot \sum_{\forall m,n} 
\begin{pmatrix} {v_\mathrm{x}[m,n,p]}\\  {v_\mathrm{y}[m,n,p]}\end{pmatrix}
\end{equation}
representing the averaged motion between two succeeding frames within the considered volume.
The maximum of the weighting function is shifted in the direction the averaged motion vector $\left( \bar{v}_\mathrm{x}[p] \hspace{4pt} \bar{v}_\mathrm{y}[p] \right)^\mathrm{T}$ is pointing. By this, it is ensured that the weighting function emphasizes the same image region over time. The overall shape of the weighting function decaying in spatial and temporal direction stays thereby unchanged. Following this description the motion compensated weighting function can be given as
\begin{equation}
\tilde{w}[m,n,p]= \begin{cases}
\tilde{\rho}[m,n,p] &, (m,n,p) \in \cal{A}\\
\delta \tilde{\rho}[m,n,p] &, (m,n,p) \in \cal{R}\\
0 &, (m,n,p) \in \cal{B}
\end{cases}
\label{equ:weighting_new}
\end{equation}
with 
\begin{equation}
\tilde{\rho}[m,n,p] = \hat{\rho}^{\sqrt{\left(m-\frac{M-1}{2}-\bar{v}_\mathrm{x}[p]\right)^2+\left(n-\frac{N-1}{2}-\bar{v}_\mathrm{y}[p]\right)^2+\left(p-\frac{P-1}{2}\right)^2}} 
\end{equation}
where $\bar{v}_\mathrm{x}[p]$ and $\bar{v}_\mathrm{y}[p]$ are the horizontal and the vertical component of the averaged motion vector of the according frame $p$ in relation to the current frame, respectively.
\vspace{-0.3cm}
\section{Simulations and Analysis}
\label{sec:simulations}
\begin{table*}[ht!]
	\begin{center}	
		\caption{\sc{Comparison of the Achieved PSNR-Values of the Simulation Results.}}
		\label{tab:results}
		\begin{tabular}{|l|c|c|c|c|}
			\hline
			\textbf{Sequence} 			& \textbf{2D-FSE \cite{seiler2010complex}}	& \textbf{3D-GF \cite{Garcia2010, Wang2012}}	& \textbf{3D-FSE \cite{seiler2016optimized}}	& \textbf{3D-MCW-FSE }\\ \hline
			{Basketball Pass} 			&	30.18 dB				& 29.70 dB				& {31.49 dB} 				&  \textbf{31.64 dB}\\ \hline
			{Blowing Bubbles} 			& 	28.04 dB				& 28.04 dB				& {29.93 dB} 				&  \textbf{30.19 dB}\\ \hline
			{BQ Square} 				&	21.02 dB				& 22.17 dB				& {23.77 dB} 				&  \textbf{23.84 dB}\\ \hline
			{Race Horses} 				&	28.40 dB				& 27.30 dB				& {28.94 dB} 				&  \textbf{30.69 dB}\\ \hline
			{Basketball Drill} 			&	31.55 dB				& 29.81 dB				& {31.27 dB} 				&  \textbf{31.80 dB}\\ \hline
			{BQ Mall} 					&	28.72 dB				& 27.12 dB				& {29.24 dB} 				&  \textbf{29.72 dB}\\ \hline
			{Party Scene} 				&	23.74 dB				& 24.13 dB				& {26.26 dB} 				&  \textbf{26.38 dB}\\ \hline
			{Race Horses} 				&	28.58 dB				& 27.09 dB				& {28.97 dB} 				&  \textbf{30.62 dB}\\ \hhline{|=|=|=|=|=|}	
			\textbf{Average}			&	27.53 dB				& 26.92 dB 	 			& {28.73 dB}				&  \textbf{29.36 dB}\\ \hline
		\end{tabular}
	\end{center}
	
\end{table*}

\begin{figure*}[t]
	\centering
	\def\svgwidth{0.77\textwidth}
	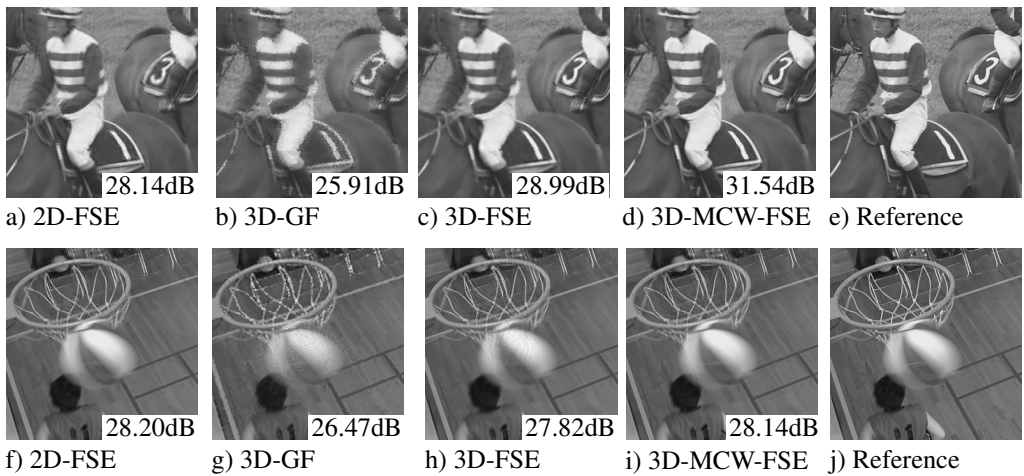
	\centering\caption{Comparison of results of the 2D-FSE, 3D-GF, 3D-FSE and 3D-MCW-FSE for the sequences Race Horses (top) and Basketball Drill (bottom).}
	\label{fig:fse_mcwfse}
\end{figure*}

For testing the proposed method, the classes C and D of the HEVC test-set \cite{bossen2013common} are used. These classes consist of four sequences each with a resolution of $832 \times 480$ and $416 \times 240$ pixels, respectively. The first fifty frames of each sequence are considered. The 3D-FSE incorporating the novel motion compensated weighting function (3D-MCW-FSE) is compared to the common 3D-FSE as presented in \cite{seiler2016optimized}. Furthermore, the presented extension of the 3D-FSE algorithm is compared to the 3D-Gap Filling algorithm (3D-GF) by Garcia et al. \cite{Garcia2010, Wang2012}. Moreover these three dimensional algorithms are compared to the 2D-FSE \cite{seiler2010complex}. For the 2D-FSE, 3D-FSE and the 3D-MCW-FSE the same parameters are used: A block-size of $4 \times 4$ respectively $4 \times 4 \times 1$, a border width of 14 and FFT of size $32 \times 32 \times 32$ is used. The weighting functions decay with $\hat{\rho}=0.7$, $\gamma$ and $\delta$ are set to $0.5$.

The results of the aforementioned simulations are shown in Table \ref{tab:results}. The given values are the average result of three reconstructions of differently sampled versions of the same sequences. As seen there, the 3D-MCW-FSE is in all cases the best performing algorithm. The highest gain can be achieved for sequences that contain much motion, as for example the Race Horses or Basketball Drill sequence. For sequences, that contain only little motion the 3D-FSE approaches the 3D-MCW-FSE as the weighting functions are nearly identical in this case. This effect can be observed with the BQ Square sequence for example.
Figure~\ref{fig:fse_mcwfse} shows some results of the simulations for the sequences Race Horses and Basketball Drill. For better visibility enlarged excerpts are displayed here. As can be seen in Figure~\ref{fig:fse_mcwfse}~d) and i), the 3D-MCW-FSE achieves much sharper edges and is capable of reconstructing finer details than the other algorithms.  
Comparing Figure~\ref{fig:fse_mcwfse} h) and i) one can observe, that the ball is distorted with structured artifacts in h). These artifacts are caused by the motion within the sequence, as the net, which is visible in previous and succeeding frames, contributes to the model up to a certain degree. This leads to a visual noticeable degradation of the image quality. Using the 3D-MCW-FSE with the motion compensated spatial weighting function, proposed here, results in a noticeable better reconstruction quality.
\noindent
\section{Conclusion and Outlook}
\label{sec:conclusion}
In this paper, the novel three-dimensional frequency selective extrapolation with motion compensated spatial weighting (3D-MCW-FSE) was presented. For video data that was recorded using a non-regular sampling sensor the proposed 3D-MCW-FSE was able to achieve a visually noticeable gain of up to 1.75 dB over the existing 3D-FSE. This gain is achieved by incorporating motion data that is obtained using optical flow.

Focus of further research will be to incorporate a scene change detection algorithm in addition, to prevent negative effects due to image content from neighboring scenes within a sequence.
\vspace{-0.3cm}

% References should be produced using the bibtex program from suitable
% BiBTeX files (here: strings, refs, manuals). The IEEEbib.bst bibliography
% style file from IEEE produces unsorted bibliography list.
% -------------------------------------------------------------------------
\bibliographystyle{IEEEbib}
\bibliography{references.bib}

\end{document}

%% file: Graphics/principle.pdf_tex
%% Creator: Inkscape inkscape 0.91, www.inkscape.org
%% PDF/EPS/PS + LaTeX output extension by Johan Engelen, 2010
%% Accompanies image file 'principle.pdf' (pdf, eps, ps)
%%
%% To include the image in your LaTeX document, write
%%   \input{<filename>.pdf_tex}
%%  instead of
%%   \includegraphics{<filename>.pdf}
%% To scale the image, write
%%   \def\svgwidth{<desired width>}
%%   \input{<filename>.pdf_tex}
%%  instead of
%%   \includegraphics[width=<desired width>]{<filename>.pdf}
%%
%% Images with a different path to the parent latex file can
%% be accessed with the `import' package (which may need to be
%% installed) using
%%   \usepackage{import}
%% in the preamble, and then including the image with
%%   \import{<path to file>}{<filename>.pdf_tex}
%% Alternatively, one can specify
%%   \graphicspath{{<path to file>/}}
%% 
%% For more information, please see info/svg-inkscape on CTAN:
%%   http://tug.ctan.org/tex-archive/info/svg-inkscape
%%
\begingroup%
  \makeatletter%
  \providecommand\color[2][]{%
    \errmessage{(Inkscape) Color is used for the text in Inkscape, but the package 'color.sty' is not loaded}%
    \renewcommand\color[2][]{}%
  }%
  \providecommand\transparent[1]{%
    \errmessage{(Inkscape) Transparency is used (non-zero) for the text in Inkscape, but the package 'transparent.sty' is not loaded}%
    \renewcommand\transparent[1]{}%
  }%
  \providecommand\rotatebox[2]{#2}%
  \ifx\svgwidth\undefined%
    \setlength{\unitlength}{225.88561051bp}%
    \ifx\svgscale\undefined%
      \relax%
    \else%
      \setlength{\unitlength}{\unitlength * \real{\svgscale}}%
    \fi%
  \else%
    \setlength{\unitlength}{\svgwidth}%
  \fi%
  \global\let\svgwidth\undefined%
  \global\let\svgscale\undefined%
  \makeatother%
  \begin{picture}(1,0.38664965)%
    \put(0,0){\includegraphics[width=\unitlength,page=1]{Graphics/principle.pdf}}%
    \put(0.35548299,0.21804998){\color[rgb]{0,0,0}\makebox(0,0)[lb]{\smash{Reconstruction}}}%
    \put(0,0){\includegraphics[width=\unitlength,page=2]{Graphics/principle.pdf}}%
  \end{picture}%
\endgroup%

%% file: Graphics/Folge_maskierter_Frames_grau_schwarz_neu_v2.pdf_tex
%% Creator: Inkscape inkscape 0.92.2, www.inkscape.org
%% PDF/EPS/PS + LaTeX output extension by Johan Engelen, 2010
%% Accompanies image file 'Folge_maskierter_Frames_grau_schwarz_neu_v2.pdf' (pdf, eps, ps)
%%
%% To include the image in your LaTeX document, write
%%   \input{<filename>.pdf_tex}
%%  instead of
%%   \includegraphics{<filename>.pdf}
%% To scale the image, write
%%   \def\svgwidth{<desired width>}
%%   \input{<filename>.pdf_tex}
%%  instead of
%%   \includegraphics[width=<desired width>]{<filename>.pdf}
%%
%% Images with a different path to the parent latex file can
%% be accessed with the `import' package (which may need to be
%% installed) using
%%   \usepackage{import}
%% in the preamble, and then including the image with
%%   \import{<path to file>}{<filename>.pdf_tex}
%% Alternatively, one can specify
%%   \graphicspath{{<path to file>/}}
%% 
%% For more information, please see info/svg-inkscape on CTAN:
%%   http://tug.ctan.org/tex-archive/info/svg-inkscape
%%
\begingroup%
  \makeatletter%
  \providecommand\color[2][]{%
    \errmessage{(Inkscape) Color is used for the text in Inkscape, but the package 'color.sty' is not loaded}%
    \renewcommand\color[2][]{}%
  }%
  \providecommand\transparent[1]{%
    \errmessage{(Inkscape) Transparency is used (non-zero) for the text in Inkscape, but the package 'transparent.sty' is not loaded}%
    \renewcommand\transparent[1]{}%
  }%
  \providecommand\rotatebox[2]{#2}%
  \ifx\svgwidth\undefined%
    \setlength{\unitlength}{225.73550007bp}%
    \ifx\svgscale\undefined%
      \relax%
    \else%
      \setlength{\unitlength}{\unitlength * \real{\svgscale}}%
    \fi%
  \else%
    \setlength{\unitlength}{\svgwidth}%
  \fi%
  \global\let\svgwidth\undefined%
  \global\let\svgscale\undefined%
  \makeatother%
  \begin{picture}(1,0.77543802)%
    \put(0,0){\includegraphics[width=\unitlength,page=1]{Graphics/Folge_maskierter_Frames_grau_schwarz_neu_v2.pdf}}%
    \put(0.06661817,0.6403295){\color[rgb]{0,0,0}\makebox(0,0)[lb]{\smash{n}}}%
    \put(-0.00386984,0.5267805){\color[rgb]{0,0,0}\makebox(0,0)[lb]{\smash{m}}}%
    \put(0.08164077,0.57311069){\color[rgb]{0,0,0}\makebox(0,0)[lb]{\smash{p}}}%
    \put(0,0){\includegraphics[width=\unitlength,page=2]{Graphics/Folge_maskierter_Frames_grau_schwarz_neu_v2.pdf}}%
    \put(0.20469271,0.04495441){\color[rgb]{0,0,0}\makebox(0,0)[lb]{\smash{Reconstructed area $\mathcal{R}$}}}%
    \put(0,0){\includegraphics[width=\unitlength,page=3]{Graphics/Folge_maskierter_Frames_grau_schwarz_neu_v2.pdf}}%
    \put(0.20469271,0.00157948){\color[rgb]{0,0,0}\makebox(0,0)[lb]{\smash{Loss area $\mathcal{B}$}}}%
    \put(0,0){\includegraphics[width=\unitlength,page=4]{Graphics/Folge_maskierter_Frames_grau_schwarz_neu_v2.pdf}}%
    \put(0.20815543,0.13283726){\color[rgb]{0,0,0}\makebox(0,0)[lb]{\smash{Volume to be reconstructed}}}%
    \put(0,0){\includegraphics[width=\unitlength,page=5]{Graphics/Folge_maskierter_Frames_grau_schwarz_neu_v2.pdf}}%
    \put(0.20577372,0.09218858){\color[rgb]{0,0,0}\makebox(0,0)[lb]{\smash{Support area $\mathcal{A}$}}}%
    \put(0,0){\includegraphics[width=\unitlength,page=6]{Graphics/Folge_maskierter_Frames_grau_schwarz_neu_v2.pdf}}%
  \end{picture}%
\endgroup%

%% file: Graphics/overview_FSE_top_2910.pdf_tex
%% Creator: Inkscape inkscape 0.92.2, www.inkscape.org
%% PDF/EPS/PS + LaTeX output extension by Johan Engelen, 2010
%% Accompanies image file 'overview_FSE_top_2910.pdf' (pdf, eps, ps)
%%
%% To include the image in your LaTeX document, write
%%   \input{<filename>.pdf_tex}
%%  instead of
%%   \includegraphics{<filename>.pdf}
%% To scale the image, write
%%   \def\svgwidth{<desired width>}
%%   \input{<filename>.pdf_tex}
%%  instead of
%%   \includegraphics[width=<desired width>]{<filename>.pdf}
%%
%% Images with a different path to the parent latex file can
%% be accessed with the `import' package (which may need to be
%% installed) using
%%   \usepackage{import}
%% in the preamble, and then including the image with
%%   \import{<path to file>}{<filename>.pdf_tex}
%% Alternatively, one can specify
%%   \graphicspath{{<path to file>/}}
%% 
%% For more information, please see info/svg-inkscape on CTAN:
%%   http://tug.ctan.org/tex-archive/info/svg-inkscape
%%
\begingroup%
  \makeatletter%
  \providecommand\color[2][]{%
    \errmessage{(Inkscape) Color is used for the text in Inkscape, but the package 'color.sty' is not loaded}%
    \renewcommand\color[2][]{}%
  }%
  \providecommand\transparent[1]{%
    \errmessage{(Inkscape) Transparency is used (non-zero) for the text in Inkscape, but the package 'transparent.sty' is not loaded}%
    \renewcommand\transparent[1]{}%
  }%
  \providecommand\rotatebox[2]{#2}%
  \ifx\svgwidth\undefined%
    \setlength{\unitlength}{226.77165354bp}%
    \ifx\svgscale\undefined%
      \relax%
    \else%
      \setlength{\unitlength}{\unitlength * \real{\svgscale}}%
    \fi%
  \else%
    \setlength{\unitlength}{\svgwidth}%
  \fi%
  \global\let\svgwidth\undefined%
  \global\let\svgscale\undefined%
  \makeatother%
  \begin{picture}(1,1.125)%
    \put(0.00893463,0.88030824){\color[rgb]{0,0,0}\makebox(0,0)[lt]{\begin{minipage}{0.16467161\unitlength}\raggedright $\cdots$\end{minipage}}}%
    \put(0.93767834,0.885961){\color[rgb]{0,0,0}\makebox(0,0)[lt]{\begin{minipage}{0.16467159\unitlength}\raggedright $\cdots$\end{minipage}}}%
    \put(0.00893146,0.64175381){\color[rgb]{0,0,0}\makebox(0,0)[lt]{\begin{minipage}{0.16467162\unitlength}\raggedright $\cdots$\end{minipage}}}%
    \put(0.93767259,0.64316701){\color[rgb]{0,0,0}\makebox(0,0)[lt]{\begin{minipage}{0.16467167\unitlength}\raggedright $\cdots$\end{minipage}}}%
    \put(0,0){\includegraphics[width=\unitlength,page=1]{Graphics/overview_FSE_top_2910.pdf}}%
    \put(0.10041651,0.99542867){\color[rgb]{0,0,0}\makebox(0,0)[lb]{\smash{n}}}%
    \put(0.06235981,0.95782772){\color[rgb]{0,0,0}\makebox(0,0)[lb]{\smash{m}}}%
    \put(0,0){\includegraphics[width=\unitlength,page=2]{Graphics/overview_FSE_top_2910.pdf}}%
    \put(0.40940541,0.99542867){\color[rgb]{0,0,0}\makebox(0,0)[lb]{\smash{n}}}%
    \put(0.37134872,0.95782772){\color[rgb]{0,0,0}\makebox(0,0)[lb]{\smash{m}}}%
    \put(0,0){\includegraphics[width=\unitlength,page=3]{Graphics/overview_FSE_top_2910.pdf}}%
    \put(0.71839435,0.99542867){\color[rgb]{0,0,0}\makebox(0,0)[lb]{\smash{n}}}%
    \put(0.68033765,0.95782772){\color[rgb]{0,0,0}\makebox(0,0)[lb]{\smash{m}}}%
    \put(0,0){\includegraphics[width=\unitlength,page=4]{Graphics/overview_FSE_top_2910.pdf}}%
    \put(0.49047258,1.0718323){\color[rgb]{0,0,0}\makebox(0,0)[lb]{\smash{ p}}}%
    \put(0,0){\includegraphics[width=\unitlength,page=5]{Graphics/overview_FSE_top_2910.pdf}}%
  \end{picture}%
\endgroup%

%% file: Graphics/vergleich_FSE_MCW-FSE_new_extended.pdf_tex
%% Creator: Inkscape inkscape 0.92.2, www.inkscape.org
%% PDF/EPS/PS + LaTeX output extension by Johan Engelen, 2010
%% Accompanies image file 'vergleich_FSE_MCW-FSE_new_extended.pdf' (pdf, eps, ps)
%%
%% To include the image in your LaTeX document, write
%%   \input{<filename>.pdf_tex}
%%  instead of
%%   \includegraphics{<filename>.pdf}
%% To scale the image, write
%%   \def\svgwidth{<desired width>}
%%   \input{<filename>.pdf_tex}
%%  instead of
%%   \includegraphics[width=<desired width>]{<filename>.pdf}
%%
%% Images with a different path to the parent latex file can
%% be accessed with the `import' package (which may need to be
%% installed) using
%%   \usepackage{import}
%% in the preamble, and then including the image with
%%   \import{<path to file>}{<filename>.pdf_tex}
%% Alternatively, one can specify
%%   \graphicspath{{<path to file>/}}
%% 
%% For more information, please see info/svg-inkscape on CTAN:
%%   http://tug.ctan.org/tex-archive/info/svg-inkscape
%%
\begingroup%
  \makeatletter%
  \providecommand\color[2][]{%
    \errmessage{(Inkscape) Color is used for the text in Inkscape, but the package 'color.sty' is not loaded}%
    \renewcommand\color[2][]{}%
  }%
  \providecommand\transparent[1]{%
    \errmessage{(Inkscape) Transparency is used (non-zero) for the text in Inkscape, but the package 'transparent.sty' is not loaded}%
    \renewcommand\transparent[1]{}%
  }%
  \providecommand\rotatebox[2]{#2}%
  \ifx\svgwidth\undefined%
    \setlength{\unitlength}{388.73810956bp}%
    \ifx\svgscale\undefined%
      \relax%
    \else%
      \setlength{\unitlength}{\unitlength * \real{\svgscale}}%
    \fi%
  \else%
    \setlength{\unitlength}{\svgwidth}%
  \fi%
  \global\let\svgwidth\undefined%
  \global\let\svgscale\undefined%
  \makeatother%
  \begin{picture}(1,0.4715017)%
    \put(0,0){\includegraphics[width=\unitlength,page=1]{Graphics/vergleich_FSE_MCW-FSE_new_extended.pdf}}%
    \put(0.40671196,0.25319032){\color[rgb]{0,0,0}\makebox(0,0)[lb]{\smash{c) 3D-FSE}}}%
    \put(0.60545572,0.25319032){\color[rgb]{0,0,0}\makebox(0,0)[lb]{\smash{d) 3D-MCW-FSE}}}%
    \put(0.80588666,0.25319032){\color[rgb]{0,0,0}\makebox(0,0)[lb]{\smash{e) Reference}}}%
    \put(0.41386563,0.01694275){\color[rgb]{0,0,0}\makebox(0,0)[lb]{\smash{h) 3D-FSE}}}%
    \put(0.60831843,0.01596302){\color[rgb]{0,0,0}\makebox(0,0)[lb]{\smash{i) 3D-MCW-FSE}}}%
    \put(0.8065118,0.01694275){\color[rgb]{0,0,0}\makebox(0,0)[lb]{\smash{j) Reference}}}%
    \put(0,0){\includegraphics[width=\unitlength,page=2]{Graphics/vergleich_FSE_MCW-FSE_new_extended.pdf}}%
    \put(0.50218483,0.28262224){\color[rgb]{0,0,0}\makebox(0,0)[lb]{\smash{28.99dB}}}%
    \put(0,0){\includegraphics[width=\unitlength,page=3]{Graphics/vergleich_FSE_MCW-FSE_new_extended.pdf}}%
    \put(0.70571124,0.28262224){\color[rgb]{0,0,0}\makebox(0,0)[lb]{\smash{31.54dB}}}%
    \put(0,0){\includegraphics[width=\unitlength,page=4]{Graphics/vergleich_FSE_MCW-FSE_new_extended.pdf}}%
    \put(0.70884628,0.04801698){\color[rgb]{0,0,0}\makebox(0,0)[lb]{\smash{28.14dB}}}%
    \put(0,0){\includegraphics[width=\unitlength,page=5]{Graphics/vergleich_FSE_MCW-FSE_new_extended.pdf}}%
    \put(0.51102966,0.04801698){\color[rgb]{0,0,0}\makebox(0,0)[lb]{\smash{27.82dB}}}%
    \put(0,0){\includegraphics[width=\unitlength,page=6]{Graphics/vergleich_FSE_MCW-FSE_new_extended.pdf}}%
    \put(0.00692881,0.01694275){\color[rgb]{0,0,0}\makebox(0,0)[lt]{\smash{f) 2D-FSE}}}%
    \put(0.20665341,0.01694275){\color[rgb]{0,0,0}\makebox(0,0)[lt]{\smash{g) 3D-GF}}}%
    \put(0,0){\includegraphics[width=\unitlength,page=7]{Graphics/vergleich_FSE_MCW-FSE_new_extended.pdf}}%
    \put(0.21074627,0.25319032){\color[rgb]{0,0,0}\makebox(0,0)[lt]{\smash{b) 3D-GF}}}%
    \put(0.00726263,0.25319032){\color[rgb]{0,0,0}\makebox(0,0)[lt]{\smash{a) 2D-FSE}}}%
    \put(0.23265739,0.38363474){\color[rgb]{0,0,0}\makebox(0,0)[lt]{\begin{minipage}{0.05187931\unitlength}\raggedright \end{minipage}}}%
    \put(0,0){\includegraphics[width=\unitlength,page=8]{Graphics/vergleich_FSE_MCW-FSE_new_extended.pdf}}%
    \put(0.10409279,0.04801698){\color[rgb]{0,0,0}\makebox(0,0)[lb]{\smash{28.20dB}}}%
    \put(0,0){\includegraphics[width=\unitlength,page=9]{Graphics/vergleich_FSE_MCW-FSE_new_extended.pdf}}%
    \put(0.30381738,0.04801698){\color[rgb]{0,0,0}\makebox(0,0)[lb]{\smash{26.47dB}}}%
    \put(0,0){\includegraphics[width=\unitlength,page=10]{Graphics/vergleich_FSE_MCW-FSE_new_extended.pdf}}%
    \put(0.10378406,0.28262224){\color[rgb]{0,0,0}\makebox(0,0)[lb]{\smash{28.14dB}}}%
    \put(0,0){\includegraphics[width=\unitlength,page=11]{Graphics/vergleich_FSE_MCW-FSE_new_extended.pdf}}%
    \put(0.30726772,0.28262224){\color[rgb]{0,0,0}\makebox(0,0)[lb]{\smash{25.91dB}}}%
  \end{picture}%
\endgroup%

%% file: Motion_adapted_3D_FSE_ICASSP_arxiv.bbl
\begin{thebibliography}{10}

\bibitem{Park2003}
S.~C. Park, M.~K. Park, and M.~G. Kang,
\newblock ``Super-resolution image reconstruction: a technical overview,''
\newblock in {\em IEEE Signal Processing Magazine}, May 2003, vol.~20, pp.
  21--36.

\bibitem{Schoeberl2011}
M.~Sch\"oberl, J.~Seiler, S.~Foessel, and A.~Kaup,
\newblock ``Increasing imaging resolution by covering your sensor,''
\newblock in {\em 18th IEEE International Conference on Image Processing}, Sept
  2011, pp. 1897--1900.

\bibitem{seiler2010complex}
J.~Seiler and A.~Kaup,
\newblock ``Complex-valued frequency selective extrapolation for fast image and
  video signal extrapolation,''
\newblock in {\em IEEE Signal Processing Letters}, Nov 2010, vol.~17, pp.
  949--952.

\bibitem{Meisinger2007a}
Katrin Meisinger and Andr{\'e} Kaup,
\newblock ``Spatiotemporal selective extrapolation for 3{D} signals and its
  applications in video communications,''
\newblock {\em IEEE Transactions on Image Processing}, vol. 16, no. 9, pp.
  2348--2360, 2007.

\bibitem{Hennenfent2007}
Gilles Hennenfent and Felix~J Herrmann,
\newblock ``Irregular sampling--from aliasing to noise,''
\newblock in {\em 69th EAGE Conference and Exhibition incorporating SPE EUROPEC
  2007}, 2007.

\bibitem{Maeda2009}
Yui Maeda and Junichi Akita,
\newblock ``A {CMOS} image sensor with pseudorandom pixel placement for clear
  imaging,''
\newblock in {\em Intelligent Signal Processing and Communication Systems,
  2009. ISPACS 2009. International Symposium on}. IEEE, 2009, pp. 367--370.

\bibitem{seiler2016optimized}
J.~Seiler, S.~Sch\"oll, W.~Schnurrer, and A.~Kaup,
\newblock ``Optimized processing order for 3{D} hole filling in video sequences
  using frequency selective extrapolation,''
\newblock in {\em Picture Coding Symposium}, Dec 2016, pp. 1--5.

\bibitem{seiler2011}
J.~Seiler and A.~Kaup,
\newblock ``Motion compensated three-dimensional frequency selective
  extrapolation for improved error concealment in video communication,''
\newblock {\em Journal of Visual Communication and Image Representation}, vol.
  22, no. 3, pp. 213--225, 2011.

\bibitem{jonscher16recursive}
M.~Jonscher, K.~Jaskolka, J.~Seiler, and A.~Kaup,
\newblock ``Recursive frequency selective reconstruction of non-regularly
  sampled video data,''
\newblock in {\em Picture Coding Symposium (PCS)}, Dec 2016, pp. 1--5.

\bibitem{seiler2015resampling}
J.~Seiler, M.~Jonscher, M.~Sch{\"o}berl, and A.~Kaup,
\newblock ``Resampling images to a regular grid from a non-regular subset of
  pixel positions using frequency selective reconstruction,''
\newblock in {\em IEEE Transactions on Image Processing}, Nov 2015, vol.~24,
  pp. 4540--4555.

\bibitem{Farnebaeck2003}
Gunnar Farneb{\"a}ck,
\newblock ``Two-frame motion estimation based on polynomial expansion,''
\newblock {\em Image analysis}, pp. 363--370, 2003.

\bibitem{Garcia2010}
Damien Garcia,
\newblock ``Robust smoothing of gridded data in one and higher dimensions with
  missing values,''
\newblock {\em Computational statistics \& data analysis}, vol. 54, no. 4, pp.
  1167--1178, 2010.

\bibitem{Wang2012}
Guojie Wang, Damien Garcia, Yi~Liu, Richard De~Jeu, and A~Johannes Dolman,
\newblock ``A three-dimensional gap filling method for large geophysical
  datasets: Application to global satellite soil moisture observations,''
\newblock {\em Environmental Modelling \& Software}, vol. 30, pp. 139--142,
  2012.

\bibitem{bossen2013common}
F.~Bossen et~al.,
\newblock ``Common test conditions and software reference configurations,''
\newblock in {\em 11th Meeting: Joint Collaborative Team on Video Coding of
  ITU-T SG}, 2011, vol.~16.

\end{thebibliography}
